# Detecting extremely low frequency primordial gravitational wave by gravitational lens system

Wenshuai Liu[1]*

[1]School of Physics, Henan Normal University, Xinxiang 453007, China.
*Email: 674602871@qq.com

ABSTRACT

Primordial gravitational waves (PGWs) are predicted to origin from inflation, according to which a period of accelerated expansion exists in the very early Universe. The detection of PGWs would verify the inflationary theory and determine its energy scale. The traditional method of using B-mode polarization to detect extremely low frequency PGW faces challenges due to the contamination from dust in Milky Way. We investigated the feasibility of using gravitational lens system (GLS) with source of high redshift to detect extremely low frequency PGW. With GLS perturbed by extremely low frequency PGWs, we found that the observed time delay in GLS could strongly deviate from the theoretical one, such strong deviation is the evidence of extremely low frequency PGWs.

INTRODUCTION

In order to solve the puzzle of why our Universe is flat, homogeneous and isotropic, theory of inflation was proposed (Guth 1981). According to inflation, in the early history of Universe, there existed a stage of accelerated expansion which stretched small scale fluctuations to superhorizon scales. Inflation produced scalar perturbations growing and giving rise to large-scale structure through quantum fluctuations at early times (Grishchuk 1976, 1977; Starobinsky 1980; Linde 1982). In addition to scalar perturbation, inflation also generated tensor perturbations known as primordial gravitational waves (PGWs) in the frequency range of $10^{-18}$ Hz $- 10^{11}$ Hz today (Starobinsky 1979; Rubakov, Sazhin & Veryaskin 1982; Fabbri & Pollock 1983; Abbott & Wise 1984; Starobinskii 1985; Allen 1988; Sahni 1990). Detecting PGWs is of great significance since it would confirm the inflationary theory and determine or constrain the energy scale of inflation. When the wavelength of PGW is larger than the horizon of the Universe, the PGW is frozen and does not evolve. It would start oscillate and propagate when the horizon size is larger than its wavelength again and, on this occasion, the amplitude of PGW deceases as the Universe expands. Thus, the amplitude of PGW is larger when its frequency is lower, meaning that extremely low frequency PGWs have the largest amplitude. The primordial spectrum of amplitude of PGW defined far outside the horizon during inflation is as (Wang, Zhang & Chen 2016)

$$h(k) = \Delta_t(k) = \Delta_R(k_0) r^{\frac{1}{2}} \left(\frac{k}{k_0}\right)^{\frac{n_t}{2} + \frac{1}{4}\alpha_t \ln(\frac{k}{k_0})}$$

where $k_0$ is a pivot conformal wavenumber which corresponds to a physical wavenumber $\frac{k_0}{a(\tau_H)} = 0.05 \text{Mpc}^{-1}$, $\Delta_R^2(k_0) = 2.1 \times 10^{-9}$ is the curvature perturbation (Ragavendra & Sriramkumar 2023) and the tensor to scalar ratio $r = \Delta_t^2(k_0) / \Delta_R^2(k_0)$ with current constrains $r < 0.032$ (Tristram et al. 2022).

The current method of detecting PGWs is the B-mode polarization imprinted on cosmic microwave background (CMB) by extremely low frequency PGWs via Thomson scattering (Kamionkowski, Kosowsky & Stebbins 1997; Seljak & Zaldarriaga 1997 ). The B-mode polarization of CMB shows to be a promising way of detecting extremely low frequency PGWs. However, due to the presence of foreground dust in Milky Way, polarized emission from such dust could contaminate the signal of B-mode polarization, making this method encounter difficulties. Although no confirmative signal of B-mode polarization induced by PGWs has been detected, constraint on the upper limit for the tensor-to scalar ratio r is improved significantly. To date, the tightest limit on r is r<0.032 (Tristram et al. 2022). Thus, it's of great significance to find a new way of detection to verify the existence of PGWs. We investigated the method of using gravitational lens system to detect extremely low frequency PGWs based on the difference between observed and theoretical time delay between different image and examined the feasibility of this method.

## THE NEW WAY OF DETECTING PGWS

This new method is based on GLS. According to general relativity, a light would bend with deflection angle $\alpha = \frac{4GM}{c^2}\frac{1}{r}$ after passing by an object with mass $M$. When the object is a galaxy composed of a planar distribution of mass elements, the total deflection angle can be calculated based on the mass distribution due to the fact that the deflection angle depends on the mass $M$ linearly and this is the fundamental of strong gravitational lensing where many images of the source are formed. Light ray travels along different path to the observer, forming different image and inducing time delay due to the different path length of the deflected light ray and the different lensing potential. The time delay is shown as $\Delta T_{ab} = T_a - T_b$ where $T_a/T_b$ is the time between the emission of radiation by the source and the signal received by the observer, specifically, in the form of comoving distance, $T = \frac{D_L D_S}{c D_{LS}}[\frac{1}{2}(\vec{\theta}-\vec{\beta})^2 - \psi(\vec{\theta})]$, where $D_L$, $D_S$ and $D_{LS}$ are the comoving distance between the observer and the deflector, that between the observer and the source and that between the deflector and the source.

The above is result with the condition that there is no perturbation from gravitational wave, what happens when PGWs are present? Like the principle of LIGO, lights are also emitted from the same location and received at the same place in GLS. It is easy to show how GLS can act as a long baseline detector of PGWs.

When the light is emitted from a source with a cosmological scale distance, it will reach the observer at Earth along different path after deflected by the gravitational lens. Due to the huge separation of location where light is deflected and the presence of PGWs, light travels along different regions of metric perturbed by PGWs, thus, one light path will be stretched or squeezed much more than another path, resulting that observed time delay could deviate from theoretical time delay. The longer the wavelength, the more obvious the deviation. GLS is sensitive to wavelength comparable to the horizon of the Universe. Thus, GLS shows to be a potential long baseline detector of gravitational wave.

Using GLS to detect PGWs was first considered by Allen (1989, 1990) with the conclusion that the additional time delays between different images of a quasar induced by extremely low frequency PGW is the evidence of PGWs. A subsequent study in Frieman et al. (1994) concluded that a gravitational lens system could not probe PGWs due to the fact that time delays induced by PGW cannot be observationally distinguishable from the intrinsic time delays produced by the geometry of the gravitational lens system. Both of Allen (1989, 1990) and Frieman et al. (1994) considered an aligned source–deflector–observer configuration, the chances of which are very low in the Universe.

Recently, study of Liu (2024) showed that GLS with singular isothermal sphere lens model and a nonaligned source–deflector–observer configuration could be used as a long baseline detector of extremely low frequency PGWs. When the GLS is perturbed by extremely low frequency PGWs, the observed time delay between images could strongly deviate from that deduced from theoretical model, meaning that the deviated time delay could be the evidence of PGWs. The deviation is defined as $\kappa = \frac{\Delta T_{Theory} - \Delta T_{Observation}}{\Delta T_{Theory}}$ where $\Delta T_{Theory}$ and $\Delta T_{Observation}$ are the theoretical time delay and the observed time delay, respectively.

However, the conclusion of Liu (2024) is reached based on the result obtained by a specific direction of propagation of PGWs with a specific direction of polarization. Whether GLS could be a detector of PGWs depends on systematic research on $\kappa$ because of the fact that PGWs are isotropic statistically with respect to direction of propagation and polarization. In addition to PGWs, there exist other perturbations which contribute to deviations in time delay, known as time delay anomaly, like a precise model of the deflector potential across the images, the distribution of mass in the lens galaxy, line of sight weak lensing, substructure in the lens and gravitational microlensing, but these perturbations can be determined or constrained with independent observational data. Thus, the existence of PGWs may be confirmed if the time delay anomaly cannot be explained by these total perturbations.

## RESULTS

Calculations show that perturbations of PGWs on GLS can be negligible when $\frac{D_{LS}}{D_S}$ is sufficiently small. Thus, the lens model could be reconstructed accurately based on the images of the two sources with low redshift, breaking the mass–sheet degeneracy. If a source with high redshift also exists in such GLS, the lens model is known to the high redshift source, meaning that the lens model used in calculating the observed time delay $\Delta T_{Observation}$ can be adopted to calculate the theoretical time delay $\Delta T_{Theory}$.

The source is at 8.4Gpc equal to redshift of 6.4 with $\beta = 2\times 10^{-7}, 3\times 10^{-7}, 4\times 10^{-7}$ relative to the lens axis while the lens is located at several different redshift. Along with the expansion of Universe, amplitude of PGW decreases if the horizon size of Universe is larger than the wavelength of PGW, resulting that the lower frequency PGW has the larger amplitude. When $r = 0.001, 0.0001$, the primordial amplitude of the PGW at a frequency of $f = 10^{-18}$ Hz is $h = 1.45 \times 10^{-6}, 4.58 \times 10^{-7}$, leading to $h = 1.45 \times 10^{-7}, 4.58 \times 10^{-8}$ at present. Other parameters are same as that in Liu (2024).

PGW at a given wavelength is isotropic statistically with respect to direction of propagation and polarization, thus, amplitude of PGW depends only on the frequency $f$. The resulting $\kappa$ from GLS perturbed by PGW with a given frequency should be the value in the form of root-mean-square (rms) $\sqrt{\langle \kappa^2 \rangle}$ where $\langle \kappa^2 \rangle$ is

$$\langle \kappa^2 \rangle = \int_0^{2\pi} \frac{d\delta}{2\pi} \int_0^{\frac{\pi}{4}} \frac{4 d\psi}{\pi} \int_0^{2\pi} \frac{d\phi}{2\pi} \int_0^{\pi} \frac{\sin\theta d\theta}{2} \kappa^2$$

where $\delta$ is the initial phase, $\phi$ and $\theta$ are the direction of propagation, $\psi$ is the direction of polarization. The rms of the theoretical time delay and the observed time delay are $\sqrt{\langle \Delta T_{Theory}^2 \rangle}$ and $\sqrt{\langle \Delta T_{Observation}^2 \rangle}$ where

$$\langle \Delta T_{Theory}^2 \rangle = \int_0^{2\pi} \frac{d\delta}{2\pi} \int_0^{\frac{\pi}{4}} \frac{4 d\psi}{\pi} \int_0^{2\pi} \frac{d\phi}{2\pi} \int_0^{\pi} \frac{\sin\theta d\theta}{2} \Delta T_{Theory}^2$$

and

$$\langle \Delta T_{Observation}^2 \rangle = \int_0^{2\pi} \frac{d\delta}{2\pi} \int_0^{\frac{\pi}{4}} \frac{4 d\psi}{\pi} \int_0^{2\pi} \frac{d\phi}{2\pi} \int_0^{\pi} \frac{\sin\theta d\theta}{2} \Delta T_{Observation}^2$$

It shows that $\kappa = \dfrac{\sqrt{\langle \Delta T_{Theory}^2 \rangle} - \sqrt{\langle \Delta T_{Observation}^2 \rangle}}{\sqrt{\langle \Delta T_{Theory}^2 \rangle}}$ can describe the deviation better than $\kappa = \dfrac{\Delta T_{Theory} - \Delta T_{Observation}}{\Delta T_{Theory}}$, thus, we use $\kappa = \dfrac{\sqrt{\langle \Delta T_{Theory}^2 \rangle} - \sqrt{\langle \Delta T_{Observation}^2 \rangle}}{\sqrt{\langle \Delta T_{Theory}^2 \rangle}}$. In order to save computational time, $\delta$ and $\theta$ are divided by ten equal intervals while $\theta$ and $\psi$ are divided by five equal intervals. Figure 1 shows that the rms of observed time delay could deviate significantly from the rms of predicted theoretical value and that $\sqrt{\langle \kappa^2 \rangle}$ is still obvious even when r=0.0001, making GLS a viable way of detecting extremely low frequency PGWs.

PGW with $f = 10^{-19}$ Hz, whose wavelength is larger than the horizon of the universe, is frozen and does not evolve. Such PGW would continue to oscillate and propagate once the horizon size is larger than it. $\sqrt{\langle\kappa^2\rangle}$ calculated with wavelength beyond horizon is much smaller than that with cross horizon wavelength, meaning that PGWs with such wavelength contribute little to the deviation.

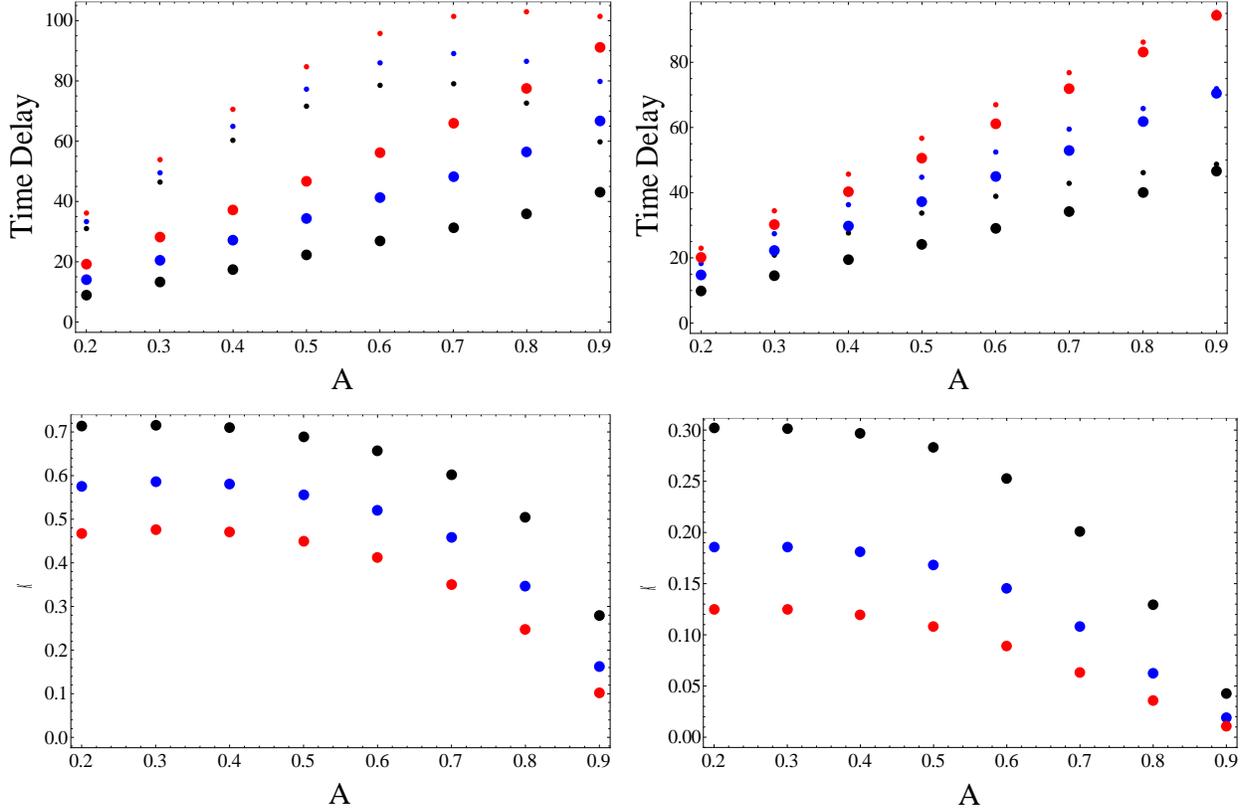

Figure 1. Top panels show the observed time delay (big dot) and the predicted time delay (small dot) with left resulting from r=0.001 and right resulting from r=0.0001. Bottom panels shows $\kappa = \dfrac{\sqrt{\langle\Delta T_{Theory}^2\rangle} - \sqrt{\langle\Delta T_{Observation}^2\rangle}}{\sqrt{\langle\Delta T_{Theory}^2\rangle}}$ with left resulting from r=0.001 and right resulting from r=0.0001. The black dot, blue dot and red dot represent the results when $\beta = 2\times10^{-7}, 3\times10^{-7}, 4\times10^{-7}$, respectively. A represents the ratio of the comoving distance between the observer and the deflector to that between the observer and the source.

## DISCUSSIONS

In order to utilize GLS as a detector of primordial gravitational waves, it is essential for the observer to confirm the existence of PGWs based on the observed time delay and the theoretical time delay. When there is only a point source in the GLS, the observer faces challenges in constructing the lens model based on the source's images to calculate the theoretical time delay due to the fact that a transformation leaves the observable images unchanged but produces a family of different mass distributions which is called the mass–sheet degeneracy. Such mass–sheet degeneracy can be broken if there are at least two point-sources at different redshifts in the GLS or an extended source. With this condition, the lens model could be reconstructed accurately using the images of the two point-sources with low redshift or the extended image of the extended source at low redshift. Simulations show that the lens model constructed from the images of two low redshift point-sources or the image of the extended source is same to that without perturbation from extremely low frequency PGWs, thus, the constructed lens model is the given lens model, meaning that the given lens model can be used to calculate not only the observed time delay but also the theoretical time delay.

The results indicate that the time delays between images can be significantly influenced by extremely low frequency PGWs and cause observable time delay deviations from the theoretical model, which suggests that GLS with SIS models could potentially serve as long baselines for detecting extremely low frequency PGWs. Therefore, in addition to B-mode polarization, the observable deviations in time delays within GLS could act as an alternative feature induced by extremely low frequency PGWs, suggesting that GLS could serve as a viable method to detect and confirm the presence of extremely low frequency PGWs. Thus, GLS provides a complementary observational feature induced by PGW besides its effect on the polarization of CMB. If, for example, the method of B-mode confirms the existence of PGWs with the upper limit constrained at r>0.001, the GLS could also detect PGWs and would verify the results.

One may regard a GLS as a rudimentary form of an interferometric gravitational wave antenna, akin to those developed in laboratory settings. In conventional laboratory instruments, the sensitivity is optimized by having the two beams of light traverse perpendicular paths. Conversely, the sensitivity of a GLS is diminished because its light rays do not adhere to perpendicular trajectories. Nevertheless, the substantial length scales involved make the GLS a detector of PGWs. In some sense, as LIGO, detecting of PGWs using GLS is a direct method.

This work explores the potential of GLS to function as moderately sensitive detectors of extremely low frequency PGWs. The sensitivity of such detectors is optimized at wavelengths that are comparable to the current Universe horizon, which corresponds to a frequency range that remains inaccessible to both laboratory and space-based instruments.

It is essential to highlight that the conclusion derived from this study is applicable not only to the specific type of deflector known as the SIS gravitational deflector but also to a broader range of lens models. Whether one adopts the particular lens model used in this research or a more generalized model, the potential existence of extremely low-frequency PGWs may be confirmed if the observed time delay significantly deviates from theoretical predictions.